\documentclass[doublecol]{epl2} 

\usepackage{amsmath,amssymb}

\newcommand \be {\begin{equation}}
\newcommand \ee {\end{equation}}
\newcommand \bea {\begin{eqnarray}}
\newcommand \eea {\end{eqnarray}}

\newcommand \ve {\varepsilon}

\title{Dissipation induced non-Gaussian energy fluctuations}
\author{Eric Bertin and Peter C.~W. Holdsworth}

\institute{Universit\'e de Lyon, Laboratoire de Physique,
\'Ecole normale sup\'erieure de Lyon, CNRS,
46 all\'ee d'Italie, F-69007 Lyon, France}

\date{\today}


\abstract{The influence of dissipation on the fluctuation statistics of the
total energy is investigated through both a phenomenological and a stochastic model for dissipative energy-transfer through a cascade of states. In equilibrium the states  obey
equipartition and the total energy obeys the central limit theorem, giving  Gaussian fluctuation.
In the presence of
dissipation the fluctuations can be driven non-Gaussian if there is macroscopic energy transfer 
 from large
to small scales. We are thus able to equate the non-Gaussian order parameter fluctuations in model equilibrium systems at criticality with energy fluctuations in these dissipative systems. 
Energy fluctuations in the phenomenological model map directly onto  the $1/f^{\alpha}$-noise problem and numerical simulations of the stochastic model yield results in qualitative agreement with these predictions.
}

\begin{document}

\maketitle

\section{Introduction}

Dissipative many body systems such as turbulent flows, granular matter or evolving dry foams
can be parameterized by spatially averaged energy functions for the integrated kinetic energy, energy dissipation, or injection in the case of driven systems. Despite the many body nature of these systems, the probability density for such global quantities is often found to be non-Gaussian (for example see \cite{Labbe,PHL,Gleeson,vanMilligen,Sandberg,Poggi01,Noullez02,Brey05,Trizac06,Brey06,Farago02,Portelli03,Farago07}), with finite skewness and a greater probability for large deviations than predicted by the central limit theorem. Although the skewness varies from one system to another, a generic pattern does seem to emerge \cite{BHP} in which the energy distribution can often be approximated by a generalized Gumbel function \cite{Bramwell00,Bramwell01,Bertin05,BC06}
--a distribution usually appearing in the context of extreme value statistics \cite{Gumbel,Galambos}. Such functions also parameterize critical order parameter fluctuations in a class of equilibrium Gaussian models sitting at their lower critical dimension.  In particular they give the form of order parameter fluctuations along the line of critical points of the 2D-XY model to an excellent approximation \cite{BHP,Bramwell00,Bramwell01}, as well as the form of one-dimensional interface fluctuations showing $1/f$ noise spectra \cite{Racz01,Racz02}. In these equilibrium systems, while the probability for order parameter fluctuations is non-Gaussian, energy fluctuations perfectly satisfy the central limit theorem, as the energy statistics for individual elements is given by equipartition.
Moving the non-Gaussian fluctuations from order parameter to energy suggests a breakdown of equipartition and the absence of a unique effective temperature covering all scales of the system. This is the subject of this paper.
In the next section, using
a model dissipative system, with energy transfer between scales and scale dependent dissipation, we show how dissipation induces a non-uniform energy profile among scales and drives the system towards a steady state with scale dependent temperature. In the third section, we test these ideas against a simple stochastic cascade model which we simulate numerically. Finally,  in the discussion, we argue that for a dissipative system with macroscopic energy transfer, the concept of equipartition of energy at equilibrium should be replaced by an equipartition of energy transfer in the steady state.


\section{A generic phenomenological model}

Our model consists of a large number of Fourier modes  $\phi_{q}$,
with $q$ a wave vector, with associated energy
$q$ is $\ve_{q}=\frac{1}{2}|\phi_{q}|^2$, considered as a random variable.
The modes with different ${q}$ values are considered to be statistically independent. 
In the spirit of a local (in scale) equilibrium approximation,
we consider that all distributions $p_{q}(\ve_{q})$ have an equilibrium form,
with a $q$-dependent temperature $T_{q}$,
\be \label{dist-epsq}
p_{q}(\ve_{q}) = \frac{1}{T_{q}} \, e^{-\ve_{q}/T_{q}},
\ee
a scenario which has been shown to be exact in a cascade model
similar in spirit to the one considered here \cite{Dauchot09}.
Note that this 'local-equilibrium' form should not be considered as a restriction to near-equilibrium situations, but rather as a mean-field-type approximation, neglecting correlations between modes.
At equilibrium, one has $T_{q}=T_0$ for all $q$, so that the total
energy $E=\sum_{q} \ve_{q}$ is simply a sum of independent and identically
distributed variables. From the central limit theorem, it follows that
the distribution of the total energy is Gaussian in the limit of a large
number of modes.
Dissipative fluid systems, such as turbulent flows and granular gases can
be modeled at this level by considering an energy cascade process from large to small length scale,
with dissipation occurring principally at small scales so that the cascade terminates at the dissipative scale $q_0$.
Note that phenomenologically, a one-dimensional cascade is often considered. In this case however, one should take into account the fact that the modes are space filling in $d$ dimensions, so that each level of the cascade consists of a shell of order $q^{d-1}$ modes $\phi_{\mathbf{q}}$. We shall come back to this point in the discussion.
For the present, we leave this issue aside
and consider a purely one-dimensional cascade.
Energy transfer between scales can result from non-local couplings between modes. However, in practical models, these couplings are often considered as relatively local in scale, as in shell models where only neighboring modes are coupled \cite{Frisch}.
Here, we adopt this viewpoint and consider for simplicity a diffusive
energy flux driven by a temperature gradient between scales,
$J_\mathrm{dif}(q) = -D(q) \partial T/\partial q$,
where $D(q)$ is a scale-dependent diffusion coefficient.
Qualitatively, $D(q)$ encodes in a mean-field spirit the non-linearities of realistic couplings.
Note that we have implicitly replaced the discrete set of temperatures
$T_{q}$ by a function $T(q)$ of the continuous variable $q$
defined on the interval $1 \le q \le q_0$.
We wish to model a system with scale-invariant dynamics,
which imposes a power-law form for the diffusion coefficient,
\be \label{eq:diffcoef}
D(q) = D_0\, q^{1+\alpha}.
\ee
For $\alpha=0$, the diffusive flux reduces to $J_\mathrm{dif}(q) =
-\partial T/\partial \ln q$, so that the dynamics is exactly the same
at all scales. The parameter $\alpha$ thus quantifies the dynamical bias
between large and small scales.
Injection takes place at $q=1$, and we further assume that
dissipation is localized at $q=q_0 \gg 1$,
so that energy is locally conserved in the interval $1<q<q_0$.
The conservation equation
$\partial T/\partial t = -\partial J_{\mathrm{dif}}/\partial q$
leads to the following diffusive evolution equation
for the temperature $T(q)$:
\be \label{eq-diff-scal}
\frac{\partial T}{\partial t} =
\frac{\partial}{\partial q} \left(
D(q)\frac{\partial T}{\partial q}\right).
\ee
This equation is supplemented by the boundary conditions at $q=1$
and at $q=q_0$.
At $q=1$ the injection mechanism is modeled by assuming that it fixes
the average value of the large scale energy: $T(q=1)=T_0$.
At $q=q_0$, the flux of dissipated energy
$J_\mathrm{dis} = \nu \, T(q_0)$
is equal to the diffusive flux $J_\mathrm{dif}(q_0)$.
The dissipation coefficient $\nu$ is assumed to be much larger
than the diffusion coefficient $D(q_0)$, consistently
with the assumption that dissipation is localized at the scale
$q_0$.

We now focus on the stationary state of the model; $\partial T/\partial t=0$, so that
$J_\mathrm{dif}(q)=J_0$, independently of scale. 
This condition enforces a temperature gradient between scales in the out-of-equilibrium steady state, thus ensuring that the system cannot support a single, effective temperature when driven away from equilibrium.
Solving Eq.~(\ref{eq-diff-scal})
with boundary condition $T(q=1)=T_0$,
the temperature profile is given after integration by
\be \label{eq-prof-temp}
T(q) = \left(T_0-\frac{J_0}{\alpha D_0}\right)
+ \frac{J_0}{\alpha D_0} \, q^{-\alpha}.
\ee
Using the equality of diffusive and dissipative fluxes at $q=q_0$,
we can determine $J_0$, yielding
\be
J_0 = T_0 \left[ \frac{1}{\nu} + \frac{1-q_0^{-\alpha}}{\alpha D_0}\right]^{-1}.
\ee
The resulting value of the flux thus depends on $\alpha$. If $\alpha>0$,
given that $q_0 \gg 1$ and $D_0/\nu \ll 1$, a finite flux $J_0$ is obtained,
$J_0 \approx \alpha D_0 T_0$, while for $\alpha<0$, the flux becomes very small
$J_0 \approx |\alpha| D_0 T_0 q_0^{-|\alpha|}$.
This transition between two regimes as a function of
$\alpha$ is similar to that reported in \cite{Dauchot09}.
From Eq.~(\ref{eq-prof-temp}),  one finds that the temperature profile $T(q)$ 
behaves for $\alpha>0$ as a power law of $q$,
\be \label{eq-Tqapos}
T(q) \approx \frac{T_0}{q^{\alpha}}.
\ee
In contrast, for $\alpha<0$, the temperature profile is
\be \label{eq-Tqaneg}
T(q) \approx T_0 \left[ 1 -
\left(\frac{q}{q_0}\right)^{|\alpha|}\right]
\ee
and is thus essentially constant, and equal to $T_0$, unless $q$
is close to $q_0$ (on a logarithmic scale).
Note that these asymptotic results rely on 
$q_0^{|\alpha|}$ being large. For small $\alpha$, one thus expects
strong finite size effects. 
The temperature gradient is therefore seen to be intimately related to finite energy transfer
down the scales and hence to energy dissipation at small scale. For negative values of $\alpha$,
although the transfer becomes small for large $q_0$, it remains sufficient to establish a local equilibrium on a sufficiently long time scale.

We now turn to the fluctuations of the total energy $E=\sum_q \ve_q$,
where $\ve_q$ is distributed according to Eq.~(\ref{dist-epsq}),
with the temperature $T_q$ satisfying either Eq.~(\ref{eq-Tqapos})
or Eq.~(\ref{eq-Tqaneg}). It is clear that the energy $E$
is a sum of independent random variables with non-identical
distributions. If $\alpha<0$, the distributions of $\ve_q$ are almost the same
for all $q$, so that the central limit theorem holds
in the limit $q_0 \to \infty$.
In the opposite case $\alpha>0$, the distribution of $\ve_q$ reads
\be
p_q(\ve_q) = \kappa q^{\alpha}\, e^{-\kappa q^{\alpha} \ve_q},
\ee
with $\kappa = 1/T_0$,
and one recovers the so-called $1/f^{\alpha}$-noise problem investigated
in \cite{Racz02}.
Hence our model provides a specific realization of the $1/f^{\alpha}$-noise problem in the spatial Fourier domain, based on explicit dissipative interactions. In contrast, standard realizations of the $1/f^{\alpha}$-noise are related to time signals, and directly postulate a $1/f^{\alpha}$ spectrum, without specifying any dynamics \cite{Racz02}.
To determine the shape of the energy distribution, we introduce the
rescaled energy $x$ defined as
\be
x = \frac{E-\langle E \rangle}{\sigma_E},
\ee
where $\langle E \rangle$ and $\sigma_E$ are the empirical mean values and
standard deviations of the energy $E$ in each case.

\begin{figure}[t]
\centering\includegraphics[width=8cm,clip]{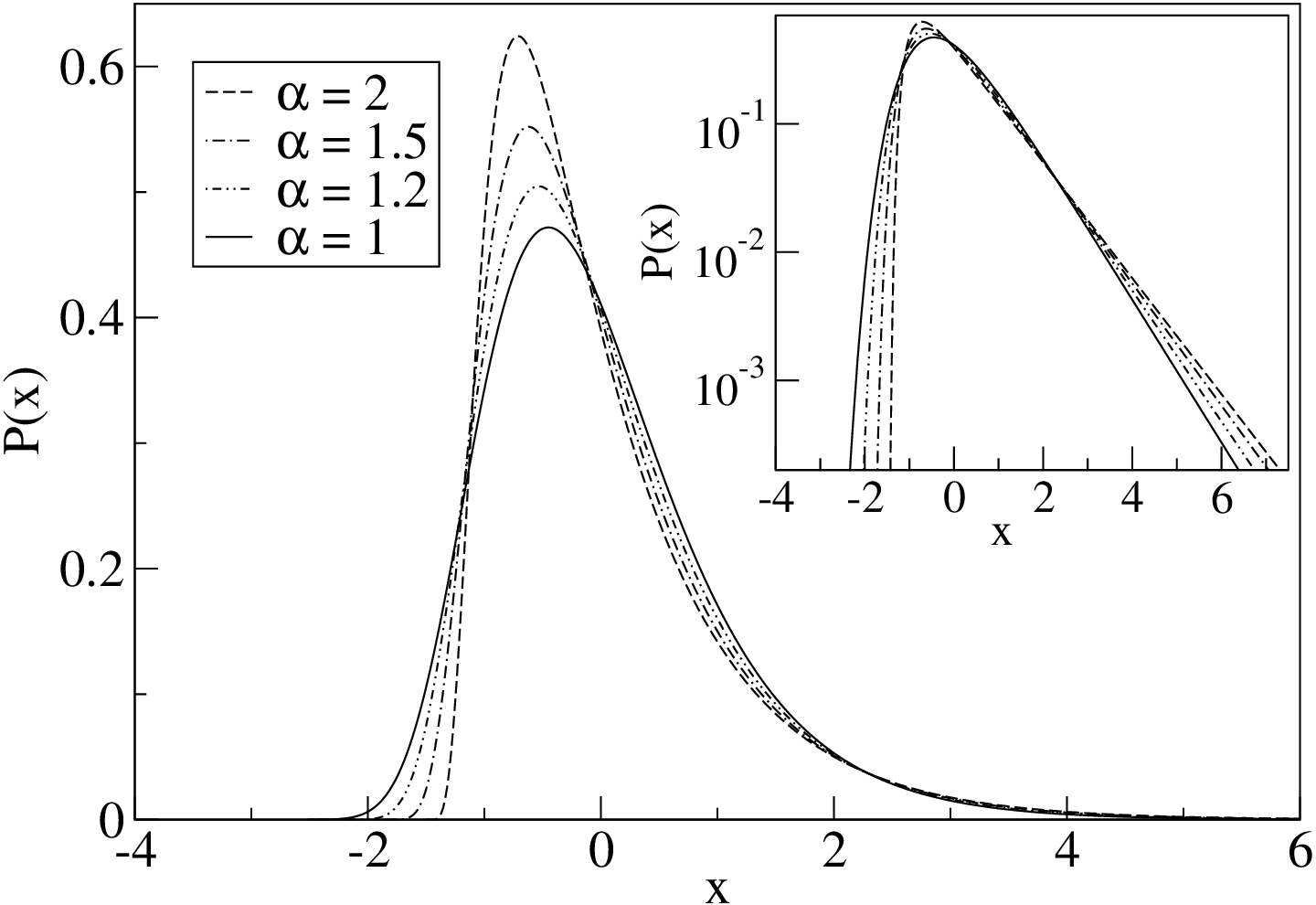}
\caption{\sl Illustration of the $1/f^{\alpha}$-noise distributions
for several values of $\alpha$.
Inset: same distributions on a semi-logarithmic scale.}
\label{fig-dist-1fa}
\end{figure}

One might expect non-Gaussian statistics to appear for all $\alpha>0$. However, 
it turns out  that the distribution $P(x)$ remains Gaussian in the large $N$ limit as long as $\alpha \le \frac{1}{2}$ \cite{Racz02}.
For $\alpha > \frac{1}{2}$, the limit distribution is non-Gaussian,
and becomes more and more asymmetric with increasing $\alpha$,
crossing over from Gaussian for $\alpha \to \frac{1}{2}^+$
to the exponential distribution for $\alpha \to \infty$.
For $\alpha=1$, one obtains the Gumbel distribution \cite{Racz01,CB08},
originally known in the context
of extreme value statistics \cite{Gumbel,Galambos}.
Remarkably, it seems that experimental and numerical 
systems often present results that can be described to a good approximation by a generalized Gumbel distribution close to this form \cite{Labbe,PHL,Gleeson,vanMilligen,Sandberg,Poggi01,Noullez02,Brey05,Trizac06}.
The non-Gaussian distributions are illustrated in Fig.~\ref{fig-dist-1fa}.
From the results of \cite{Racz02}, one easily obtains
the skewness and kurtosis of these distributions as
\bea
\langle x^3 \rangle &=& \frac{2\,\zeta(3\alpha)}{\zeta(2\alpha)^{3/2}}\\
\langle x^4 \rangle-3 &=& \frac{6\,\zeta(4\alpha)}{\zeta(2\alpha)^2}
\eea
where $\zeta$ is the Riemann Zeta function, defined as
$\zeta(a) = \sum_{n=1}^{\infty} n^{-a}$.
Note that by definition $\langle x \rangle=0$ and $\langle x^2 \rangle=1$.
Both the flatness and the kurtosis go to zero when
$\alpha \to \frac{1}{2}^+$, consistently with the fact that the $1/f^{\alpha}$-noise distributions
converge to the Gaussian law in this limit.

To sum up, the main outcome of the generic phenomenological model considered in this section is a transition
between Gaussian and non-Gaussian energy distributions,
occuring when the parameter $\alpha$ characterizing the transfer properties between scales is varied.
For a finite number of modes, the transition should be replaced by a crossover between close-to-Gaussian and strongly non-Gaussian distributions.
It is plausible that the precise crossover value of $\alpha$ may depend on the details of the model, but one may expect the scenario to be qualitatively valid in more generic situations. In the next section, we test this scenario in an explicit stochastic model.

\section{Stochastic dissipative model}

As a test of the above scenario, we consider
a dissipative energy transfer model explicitly defined by stochastic rules.
This model is similar in spirit to the ones considered in
\cite{Bertin05,Bertin06,Dauchot09},
though some significant differences are present.
The energy of mode $q$ is $\ve_q=\frac{1}{2}|\phi_q|^2$,
where  $\phi_q$ is a Fourier amplitude, attached to the wavenumbers $q=1,\ldots,q_{\max}$.
The dynamics is defined by three different mechanisms: energy injection
at large scale, internal energy transfer between neighboring scales and local
energy dissipation.
To preserve scale invariance, the presence of energy transfer between wavenumbers $q$ and $q'$ is determined according to the ratio $q'/q$; namely
it occurs only if $1/\lambda < q'/q <\lambda$, where $\lambda>1$ is a fixed parameter.
The different transfer rates are defined as follows:
an amount of energy $\mu$ is either injected 
at $q=1$ with rate (probability per unit time) $I(\mu)$, transferred from a site $q$ to
site $q'$ with a rate $\Gamma_{q,q'}(\mu|\ve_q)$, or dissipated from site $q$
at a rate $\Delta_q(\mu|\ve_q)$.
Note that the transfer and dissipation rates depend on the energy $\ve_q$
present at wavenumber $q$.
We choose the following forms for the rates
\bea
\label{eq-Jmu}
I(\mu) &=& T_0^{-1}\, e^{-\mu/T_0} \,,\\
\label{eq-phimu}
\Gamma_{q,q'}(\mu|\ve_q) &=& D_0 (qq')^{\alpha/2} \, \ve_q \, ,\\
\label{eq-deltamu}
\Delta_q(\mu|\ve_q) &=& \nu_0 q^2 \, \ve_q \,.
\eea
Choosing the form (\ref{eq-Jmu}) for $I(\mu)$ means that energy is injected
at wavenumber $q=1$ by connecting this element to a thermostat at temperature
$T_0=1$, which fixes the energy scale.
The form (\ref{eq-phimu}) of the transfer rates is chosen to be symmetric in $q$, $q'$ (so as to satisfy detailed balance), and to reproduce the scaling of the effective diffusion coefficient Eq.~(\ref{eq:diffcoef}).
Given that $q$, $q'$ are close, $(qq')^{\alpha/2} \sim q^{\alpha}$, and as the number of modes $q'$ in the shell $[q/\lambda,\lambda q]$ is $\sim q$,
the total transfer rate from wavenumber $q$ is indeed $\sim q^{1+\alpha}$, as in the phenomenological model.
Energy is also transferred back to the reservoir with a rate $\tilde{I}(\mu|\ve_q)=\ve_q$.
The value of the parameter $\alpha$ is again the key to the way in which energy is transfered
between scales.
The constant $D_0$ in Eq.~(\ref{eq-phimu}) is chosen such that
$D_0 (qq')^{\alpha/2}$ remains bounded when $q_{\max} \to \infty$.
We take $D_0=1$ for $\alpha \le 0$ and $D_0 = q_{\max}^{-\alpha}$
for $\alpha > 0$.
Finally, the constant $\nu_0$
controls the dissipation rate and can be interpreted as a viscosity.
In the following, we shall be interested in the limit where $\nu_0$ takes small,
but non-zero values.
Note that the dissipation rate is the only dynamical rule that breaks detailed
balance. In the absence of dissipation ($\nu_0=0$) the system
converges to the equilibrium Gibbs distribution with temperature $T_q=T_0$ at all scales.
We shall see however that the limit $\nu_0 \to 0$ is singular,
as is well-known in the context of fluid turbulence \cite{Frisch}.

The main differences between the present cascade model and that introduced in \cite{Bertin05}
are that in the latter the energy transfer rate is fully biased
and scale-independent and that the dissipation rate has
an ad hoc form chosen to make the model analytically solvable. Here, in addition to the inclusion of back transfer from large to small scale, we choose a dissipation rate  proportional to $q^2$, which is natural
in the context of viscous damping.

With this form, as long as $\alpha < 2$, the ratio of the dissipated to transferred energy increases as the wavenumber $q$ increases,  assuring a cut-off wavenumber for the cascade:
\begin{equation}
q_0 \sim \nu_0^{1/(\alpha-2)}D_0^{\alpha/(\alpha-2)}.
\end{equation}
In the context of simulations of a finite system with a number $q_{\max}$ of modes, the characteristics of the cascade are determined by  both $q_0$ and $q_{\max}$ and their interplay, as we show below.

\begin{figure}[t]
\centering\includegraphics[height=5cm,clip]{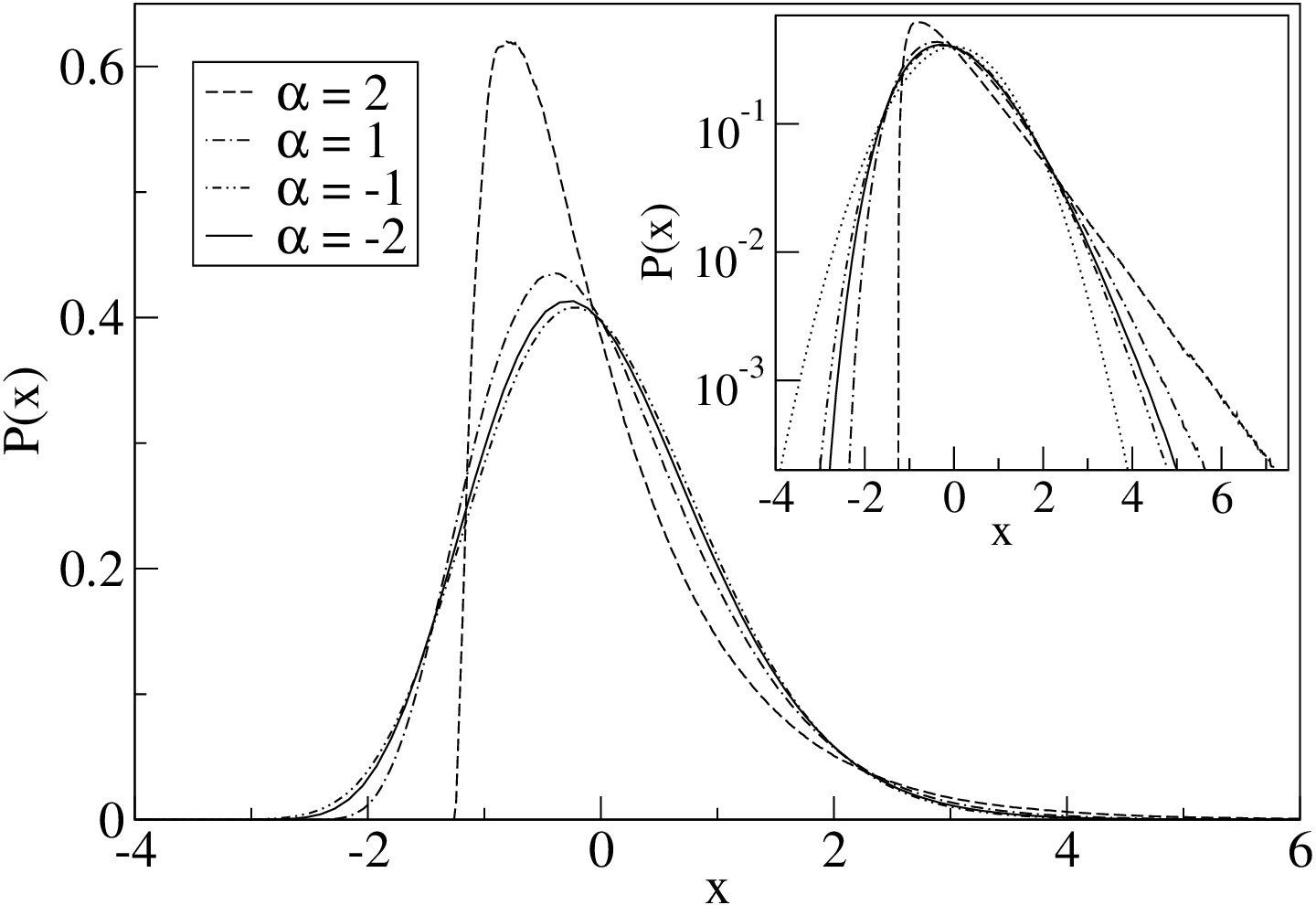}
\centering\includegraphics[height=5cm,clip]{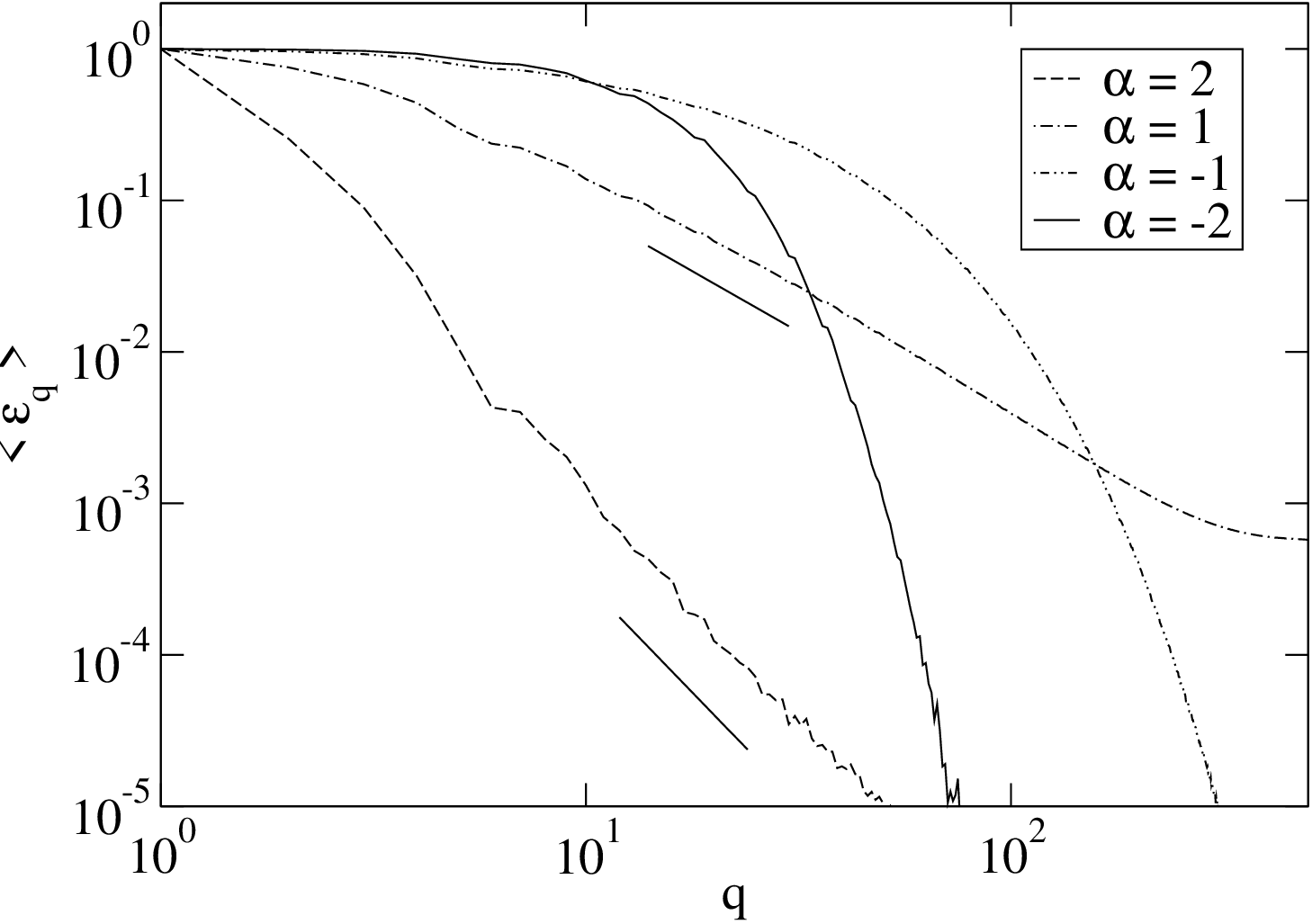}
\caption{\sl Top panel: Distribution of the rescaled energy
$x=(E-\langle E \rangle)/\sigma_E$ in the stochastic cascade model
for different values of the $\alpha$ characterizing energy transfer
through scales ($\nu_0=10^{-5}$).
Inset: same data on a semi-logarithmic scale. The dotted line is the Gaussian
distribution, shown for comparison.
Bottom panel: energy spectrum $\langle \ve_q \rangle$ for the same values of
$\alpha$, on a log-log scale. Straight lines indicate slopes of $-2.9$ ($\alpha=2$) and $-1.6$ ($\alpha=1$).
}
\label{fig-dist-en}
\end{figure}

We have simulated the model for $q_{\max}=500$, and measured the
steady-state distribution of the total energy for several values of $\alpha$.
The resulting distributions $P(x)$ of the rescaled energy $x$ are shown for $\nu_0=10^{-5}$ in the upper panel 
 Fig.~\ref{fig-dist-en}, on a linear scale in the main figure and log-linear scale in the inset.
For $\alpha=2$ the shape is strongly non-Gaussian, with a clear skewness and long tail for 
fluctuations for positive values of $x$. As $\alpha$ decreases from $2$ to $1$ there is a sharp decrease in
the anisotropy of the distribution in qualitative agreement with the data for the phenomenological model shown in 
Fig. \ref{fig-dist-1fa}. The form for $\alpha=1$ is close to that typically observed in experiments and simulations for dissipative systems (for example see \cite{Labbe,PHL,Gleeson,vanMilligen,Sandberg,Poggi01,Noullez02,Brey05,Trizac06}). Further reduction in the value of $\alpha$ sees a slow evolution towards a Gaussian distribution although it is not possible to determine clearly from this data if a change
in regime occurs for $\alpha$ around zero.

Evidence for such a crossover is however given by the evolution of the energy spectrum $\langle \ve_q \rangle$
with $\alpha$, shown in the lower panel of Fig.~\ref{fig-dist-en}. There is a clear quantitative difference between the spectra for positive and negative values of $\alpha$. For $\alpha=2$, the spectrum is approximately power law over five orders of magnitude in energy (with an exponent $\tilde{\alpha}\approx -2.9$),
with no evidence of a crossover to a dissipative regime, which is consistent with the ratio of energy transfer to dissipation being scale independent.
For $\alpha=1$ the inertial range of approximate power law scaling
(here with an exponent $\tilde{\alpha}\approx -1.6$)
is cut off by a build up of energy at a scale around $q=2\times 10^2$,  which corresponds to the dissipation scale, $q_0$ for this value of $\nu_0$. One might have expected a fall off in energy per mode as one goes to larger wavenumber.
However, in this case $q_0$ is close to $q_{\max}$ with the result that energy is stored at the bottom of the cascade, rather than being dissipated at larger wavenumber. 
Note that when simulations are run for smaller values of $\nu_0$, we find that  the measured exponents
 $\tilde{\alpha}$ decrease in magnitude (data not shown) towards $\alpha$, as for the phenomenological model presented in the previous section. 
More extensive finite size scaling is necessary here,
but  the general behavior of extensive energy transfer, inertial scaling regime and non-Gaussian energy fluctuations appears well established. 
For negative values of $\alpha$ the spectrum again mirrors that for the phenomenological model, being constant at small $q$ and cut off to zero around the characteristic dissipation scale, $q_0\sim 50$ for $\alpha=-1$ and $q_0\sim 15$ for $\alpha=-2$.

\begin{figure}[t]
\centering\includegraphics[height=5cm,clip]{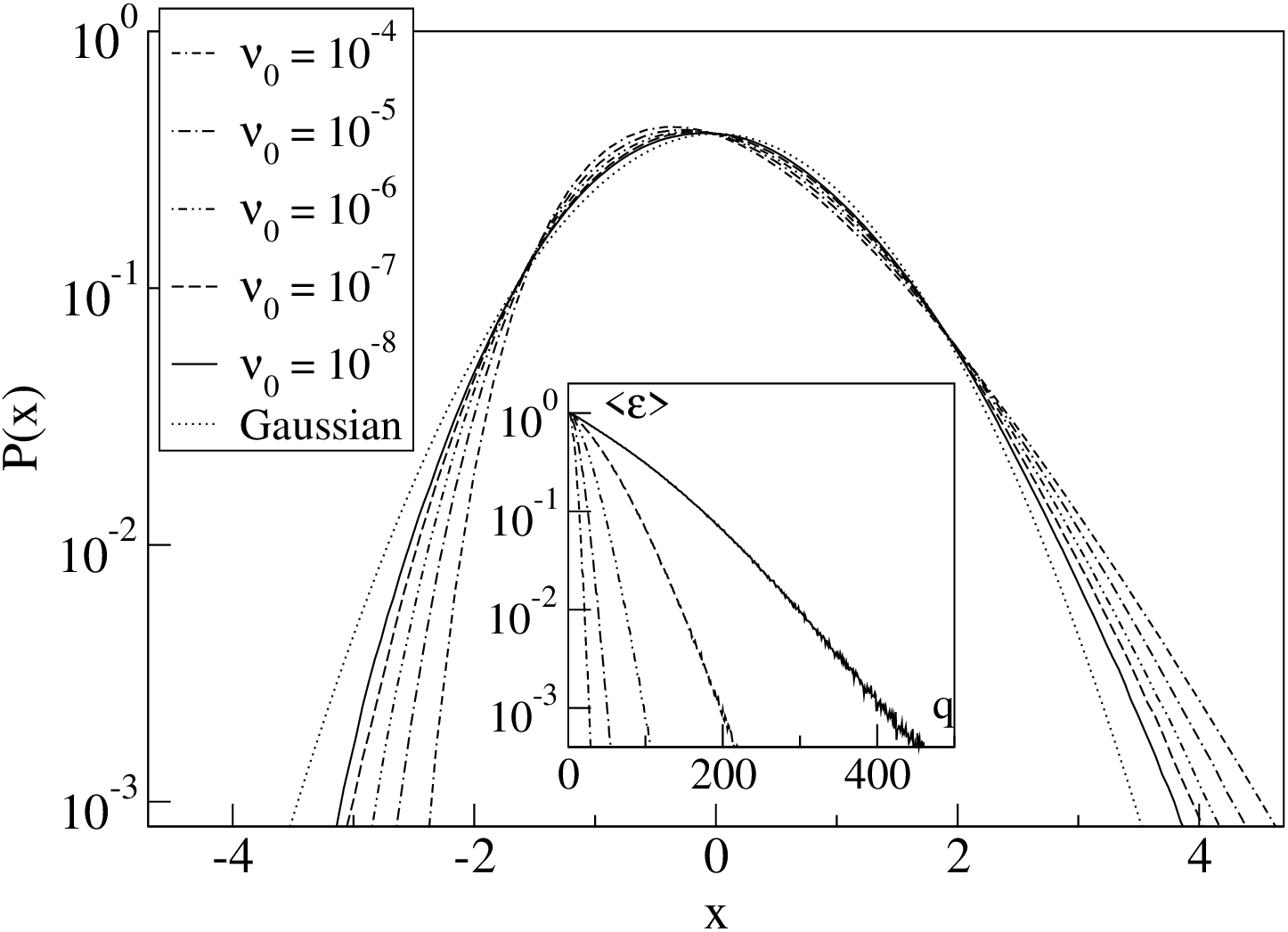}
\caption{\sl Distribution $P(x)$ of the rescaled energy $x$
for decreasing values of the viscosity $\nu_0$ ($\alpha=-2$).
A slow convergence to the Gaussian distribution (dotted line) is observed.
Inset: average energy $\langle \ve_q \rangle$ as a function of $q$
for the same values of $\nu_0$ ($\nu_0$ decreases from bottom to top). 
}
\label{fig-dist_alp-2}
\end{figure}

Deviation from a Gaussian distribution for $\alpha \lesssim 0$ can
be traced to the effects of finite viscosity, as shown in in Fig. \ref{fig-dist_alp-2}, where
we compute
$P(x)$ for decreasing values of the $\nu_0$. As in this case the crossover scale $q_0 \sim \nu_0^{-1/4}$, we are able to push $\nu_0$ as low as $10^{-8}$ before the maximum cut off $q_{\max}=500$ plays a role, allowing for a significant scaling range. 
A slow convergence to a Gaussian is observed
as $\nu_0$ decreases.
The convergence corresponds to an increasing number of equilibrated  degrees of freedom
as $q_0$ increases (see inset of Fig.~\ref{fig-dist_alp-2}) and an approach to the central limit theorem result as  $\nu_0 \to 0$.

\begin{figure}[t]
\centering\includegraphics[height=5cm,clip]{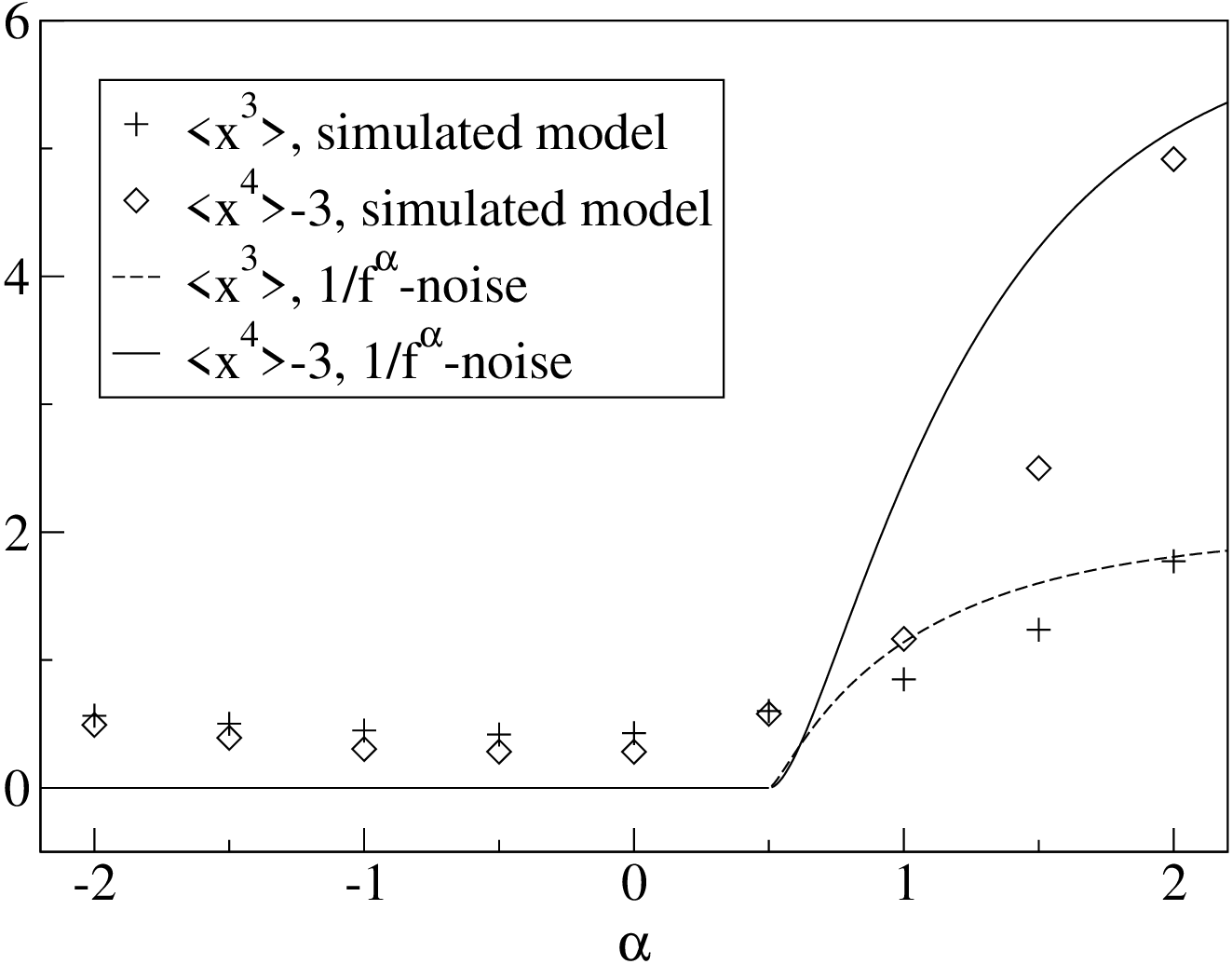}
\caption{\sl Skewness and kurtosis as a function of $\alpha$
for the simulated model (symbols) and for the $1/f^{\alpha}$-noise
distributions (lines), with a viscosity $\nu_0=10^{-5}$.}
\label{fig-moments}
\end{figure}

To provide further evidence for a crossover from Gaussian to non-Gaussian behavior
as a function of $\alpha$, we compute the skewness $\langle x^3 \rangle$
and the kurtosis $\langle x^4 \rangle-3$, 
as a function of $\alpha$ for fixed $\nu_0=10^{-5}$. Results are plotted in the Fig.~\ref{fig-moments}.
The skewness and kurtosis take small values for $\alpha \lesssim 0.5$,
and larger values for $\alpha \gtrsim 0.5$. The data can be compared with the analytic predictions from the phenomenological model. Although we are not able to make a quantitative statement, the simulation results are clearly consistent with the phenomenology within the accuracy achieved. For $\alpha \lesssim 0$, the skewness and kurtosis can be made smaller
by decreasing the viscosity $\nu_0$, as shown in Fig. \ref{fig-dist_alp-2}.
In contrast, for positive values of $\alpha$, numerical simulations are difficult for very low
viscosities, as $q_{\max}$ must exceed $q_0$ to avoid energy accumulation at large wavenumber. A more detailed study therefore requires finite size scaling both in viscosity and $q_{\max}$ which we leave for a future study.
However the results presented in this section show that the scenario of an energy transfer driven crossover between near-Gaussian and strongly non-Gaussian energy fluctuations is also present in this stochastic cascade model.

\section{Discussion}

In the previous sections we have shown that the asymmetry of the energy distribution
in dissipative systems can be traced back to the efficiency of the transfer
mechanism from large to small scales. This departure from the central limit theorem has, in the past been compared with non-Gaussian order parameter fluctuations in model systems of criticality at equilibrium \cite{BHP,Bramwell01,Racz01}. 
The asymmetry can here be understood as resulting from
the constraint imposed by the equipartition of energy.
Consider the Fourier amplitude of the order parameter for wave vector $\mathbf{q}$, $\psi_{\mathbf{q}}$ in the specific example of the low temperature phase of the 2D-XY model:
the energy associated with the mode reads $\epsilon_{\mathbf{q}}=
\frac{1}{2}\, q^2 |\psi_{\mathbf{q}}|^2$, while the contribution of wavevector ${\mathbf{q}}$
to the order parameter is $m_{\mathbf{q}} = \frac{1}{2}\, |\psi_{\mathbf{q}}|^2$
\cite{Bramwell01}.
The equipartition of energy at equilibrium imposes that the average energy $\langle \epsilon_{\mathbf{q}} \rangle$
of mode ${\mathbf{q}}$ is independent of ${\mathbf{q}}$, namely
$\langle \epsilon_{\mathbf{q}} \rangle=k_B T$.
The equipartition thus imposes $\langle m_{\mathbf{q}} \rangle = k_B T/q^2$.
Combining the $1/q^2$ scaling for $m_{\mathbf{q}}$ with the density of states, $g(q) \sim q$ results in non-Gaussian fluctuations
for the order parameter, while the fluctuations of the total energy are perfectly Gaussian, being the sum of independent and identically distributed random variables. Similar behavior occurs in one dimension if interactions are long ranged \cite{Joanny}, leading directly to a "$1/f$-noise" spectrum \cite{Racz01}, which can in turn be generalized to the $1/f^{\alpha}$ behaviour \cite{Racz02} discussed above. This is of course an extremely simplified vision of
criticality \cite{Clusel06}:
non-linear coupling between modes can drive a divergent specific heat and consequently non-Gaussian energy fluctuations. However, the specific heat exponent $\alpha$ is systematically smaller than that for the susceptibility, $\gamma$, indicating a weaker divergence and showing that energy fluctuations are in general more constrained than those for the order parameter. 

Here we consider dissipative systems at the same level of approximation, finding 
clear violation of the central limit theorem and non-Gaussian statistics for the spatially averaged
internal energy. In the context of an energy cascade, with dissipation localized at small length scales, equipartition  is replaced, in steady state, by a new constraint; the average energy transfer between scales is uniform throughout the cascade, $J_q=J_0$. For the models with diffusive dynamics that we have considered, this steady state condition ensures that global equipartition of energy must be violated, thus introducing a scale dependent temperature, $T_q\sim q^{-\alpha}$ and non-Gaussian energy fluctuations, if $\alpha$ is large enough.  Indeed, as $\epsilon_q = |\phi_q|^2$, one finds for both the phenomenological and stochastic models that $J_q \sim q^{\alpha} |\phi_q|^2$ in analogy with the energy of the equilibrium system.
Hence we find quantitative equivalence between order parameter fluctuations in the equilibrium models and those for internal energy in the case of dissipation. The sum over all instantaneous energy transfers would thus satisfy the central limit theorem.

The comparison with equilibrium systems can be pushed further by considering the role of spatial dimension, $d$ in the presence of dissipation.
At a heuristic level, we simply multiply the $1/f^{\alpha}$ spectrum by the density of states $g(q) \sim q^{d-1}$, yielding an effective exponent $\alpha'=\alpha-(d-1)$ for the energy spectrum.
If $\alpha$ is large enough, a transition from non-Gaussian to Gaussian fluctuations is observed when increasing $d$ beyond $d_{\mathrm{u}}=\alpha+\frac{1}{2}$ (corresponding to $\alpha'=\frac{1}{2}$), which can be interpreted as an upper critical dimension, in qualitative analogy with equilibrium critical phenomena.
Correlations between the modes at each level of the cascade will however reduce the number of effective degree of freedom, placing the effective exponent in the range, $\alpha-(d-1) < \alpha' <\alpha$.

Finally, one might ask: what do we learn about experimental systems or more realistic models from this work?
There are of course a barrage of simplifications separating our phenomenology from experiment, so that comparison with turbulent flow, or dense granular media at a microscopic level is not directly feasible. However, our arguments do lay out a general framework for why energy fluctuations are more violent in dissipative systems than in equilibrium. Hence we outline the context in which one can expect energy fluctuations for a system in steady state to be approximated by a $1/f^{\alpha}$-like distribution. Taking a step beyond phenomenology requires the identification of our models as the harmonic limit of more complex model systems, as was done for critical order parameter fluctuations in equilibrium \cite{Clusel06}. Particularly interesting directions to follow are the
kinetic energy fluctuations of a granular gas in a homogeneous cooling state \cite{Brey06}, or of decaying Burgers turbulence \cite{Noullez02}. In both these cases the energy decays in time in a self similar manner, allowing the decay to be rescaled away through a change of variables, leaving what could be considered as an idealized steady state. In these transformed variables
 the agreement between simulation results and the model distributions is close enough to suggest that the phenomenology presented here becomes correct in these specific cases.

In conclusion, our main results are two-fold. Firstly, we have shown that there generically exists a crossover from Gaussian to strongly non-Gaussian fluctuations  of the spatially averaged energy, when varying the transfer properties through scales (quantified here by the parameter $\alpha$) in dissipative systems. Secondly, we have argued that in the non-Gaussian regime, the asymmetry of the energy distribution is driven by a finite energy flux crossing the system from injection to dissipative scales. The constancy of this flux across scales plays  a role similar to that of energy equipartition. In consequence non-Gaussian energy fluctuations arise in complete analogy to those for the order parameter at equilibrium.
As for future work, it would be of interest to try to quantitatively validate this scenario in more realistic models such as shell models, Burgers turbulence or granular gases.

 \section{Acknowledgments} It is a pleasure to thank Steven T Bramwell, Maxime Clusel and Olivier Dauchot for useful discussions and related collaborations.

\end{document}